# Expressions and Formulas to Expand and Add Potentiality to Knowledge Underlying Diffraction

Noureddine HADJI

Département de Physique, Université Badji Mokhtar, Annaba, BP 12 Annaba 23000, Algérie
*Alternative address*: 138, Villa 60, Cité des Jardins, Annaba 23 200, EL HADDJAR, Algeria.
*Email*: noureddine.hadji@univ-annaba.dz

https://orcid.org/0000-0001-5876-8002

**Abstract:** Development generating diffraction-related valuable expressions and formulas capable of initiating new era for diffraction and for scientific domains that use it provided. The main expression, among these, gives diffracted intensity as explicit function of true and absolute distribution of interatomic vectors (DIV) and is suitable for constructing analytical techniques capable of converting experimental diffraction intensity data into real structural data, immediately and accurately interpretable as such, through Fourier transform tool (FTT) with no assumption and no approximation additional to those involved in kinematic theory. The first ever produced formula giving distribution of interatomic vectors, and hence distribution of interatomic distances (DID), of monatomic crystal provided together with description of possible way of obtaining it. Given examples of application relate to getting two DIDs and to using them to calculate X-ray diffracted intensity profiles, one DID was for zinc blende (ZnS) and was calculated from atom positions and the other one was for polonium and was calculated using both atom positions and appropriate formula. Software for calculating DIDs and X-ray profiles for monatomic crystals available freely. No longer needed to affect Debye scattering equation (DSE) through approximation and assumption makings before combining it with FTT to get to experimental structural data.

**ملخص:** تعبيرات و صيغ ثمينة مولدة للتطور مرتبطة بالانعراج قادرة على بدء عهد جديد بالنسبة للانعراج و بالنسبة للميادين العلمية التي تستعمله موفرة. التعبير الرئيسي ما بين هذه يعطي شدة الانعراج كدالة واضحة لتوزيع المتجهات ما بين الذرات (تمتم) الصحيح و المطلق و هو تعبير لائق لبناء تقنيات تحليلية قادرة على تحويل بيانات شدة الانعراج تجريبية إلى بيانات بنيوية، قابلة للتأويل على هذا النحو فورا و بدقة، حقيقية عن طريق أداة تحويل Fourier (أتف) بدون استعمال أي افتراض أو أي تقريب إضافي لتلك المتورطة في النظرية الكينماتية. أول صيغة منتجة على الإطلاق

معطية لتمتم، و بالتالي لتوزيع المساقات ما بين الذرات (تمسم)، لبلورة أحادية الذرات موفرة مع وصف لطريقة ممكنة للتحصل عليها. الأمثلة التطبيقية المعطاة تتعلق بالتحصل على اثنين من التمسم و استعملهما لحساب رسمين الانعراج جانبين بالأشعة السينية، أحد التمسم هو لمزيج الزنك (ZnS) و حصل عليه عن طريق الحسابات استعمالا المواقع الذرية و الآخر هو للبولونيوم و حصل عليه عن طريق الحسابات استعمالا كلا من المواقع الذرية و الصيغة المناسبة. برنامج الحاسوب للقيام بحسابات تمسم و رسومات الانعراج جانبية بالأشعة السينية المناسبة للبلورات الأحادية الذرات متوفرة مجانا. لم تعد هنالك حاجة لتأثير على معادلة التبعثر لـDebye (DSE) عبر اتخاذ قريبات و افتراضات قبل ضمها إلى الأتف للوصول إلى بيانات بنيوية تجريبية.

## 1. Introduction

The synthesis of line-profiles on the basis of a microstructure model and application of the (kinematical) diffraction theory without any further assumption may be the two parts of the most promising development in methods for diffraction line-profile analysis, [1]. A possible route to, at least, the second part of this statement can be through doing more work on the usual expression of the diffracted intensity to get to an equivalent expression that can be used with no further assumption. Effectively, in the kinematical diffraction theory the main expression of the diffracted intensity, for resulting from a work stopped before exhausting all possible resources, is a *double sum*, for instance see [2]-[4], and, thus, cannot be combined "as is" with the Fourier transform tool (FFT) to build analytical techniques for structural analyses without "accuracy-affecting" additional assumptions and approximations. Here, continuing with the work, we get, making absolutely no approximation and no assumption, to a considerably more advantageous main expression that is *naturally* a *single sum* and that is an explicit function of the, handy, material's atomic arrangement-related *absolute* distribution of interatomic vectors (DIV) present within the diffracting volume. Therefore, this expression allows applying the kinematical diffraction theory together with FTT using no further assumption, and no approximation, to get accurate and straightforwardly interpretable structural information. It also paves the road for producing more rapidly diffracted intensity profiles through either of DIV and distribution of interatomic distances (DID) *formulas* if available. We give here the first DIV and DID formulas for triclinic monatomic crystals ever produced and expect this would activate induction of extra progress in structural information production through diffraction for both crystals and disordered materials with valuable impacts almost in all fields of sciences involving the diffraction tool, for instance materials science, molecular biology etc. We give here, as a concrete example, a reshaped version of the Debye scattering equation (DSE), e.g. see [2]-[8] for this DSE, deduced from our main expression, which can be used with no further assumption and no approximation for constructing analytical techniques that can extract, using FFT, accurate *absolute* DIDs for disordered materials from experimental diffraction data.

The diffraction tool, as a means for structural studies, has been greatly contributing, [9]-[11]-[13] to the development of materials science and molecular biology since the discovery, [14], in 1912 of x-ray diffraction by crystals. These studies involve various techniques, for instance [2]-[4]-[6]-[8]-[11]-[15]-[22], some of which consider diffracted intensity calculations, e.g. [2]-[4]-[6]-[8], and/or conversions of experimental diffracted data into structural data through FTT, e.g. [2]-[4]-[8]-[11]-[15]-[16], others, e.g. [17], don't. The principal aim of this work is to provide new progress producing contributions within the frame of the kinematic approach to the diffraction theory, and, therefore, to also advance and enlarge further the background underlying the usually employed means for acquiring knowledge through diffraction (of x-rays, electrons, neutrons). Thus, in the framework of this aim, additional work was performed on the usual expression, given by (1) below, of the diffracted intensity by a material to get to a different but fully equivalent and much more gainful expression. Among the valuable characteristics this new expression has is the fact that it does have the form suitable for constructing analytical techniques, within the limits of validity of the kinematic approach to diffraction, making no additional approximation and no additional assumption; techniques like those that include: (i) intensity calculations either through **formulas** (very rapid indeed) or from known atom positions (rapid); and (ii) the use of FTT to extract accurate experimental real space data from reciprocal space data which are immediately and validly interpretable on the basis of no more than the physical concepts and principles the kinematic approach to diffraction relies on. However, giving descriptions of techniques for deducing structural data from reciprocal space data is not in the aims of this work but providing formulas for DIV and DID through which calculated diffracted intensities can be obtained for monatomic crystals with triclinic lattices is. Examples of calculated DIDs and intensities are given.

## 2. The expressions of the diffracted intensity

### 2.1. The so far existing expressions of the diffracted intensity

The work carried out to get the kinematic diffracted intensity expressions normally used in structural studies of materials was stopped at the level of the following equations:

1) the usual kinematic, static, expression for the intensity, $I(\vec{S})$, scattered by the $N$ atoms present within the portion of volume, of a material system, irradiated by an incident beam (X-ray, ...), e.g. [2]-[4] (for an expression of the intensity diffracted by an inhomogeneous material system see [23]):

$$I(\vec{S}) = \sum_{i=1}^{N}\sum_{j=1}^{N} f_i(\vec{S})f_j(\vec{S})\exp(-i\vec{S}.\vec{r}_{ij}) = \sum_{i=1}^{N}\sum_{j=1}^{N} f_i(\vec{S})f_j(\vec{S})\cos(\vec{S}.\vec{r}_{ij}) \quad (1)$$

[where $f_i(\vec{S})$ and $f_j(\vec{S})$ are the atomic scattering factors of, respectively, the $i$th and the $j$th atoms; $\vec{r}_{ij}$ is the vector joining the $i$ and $j$ atoms; $\vec{S}$ is the scattering vector defined as $\vec{S} = (4\pi/\lambda)\sin\theta.\vec{u}$ where $\lambda$ is incident beam wavelength, $2\theta$ is scattering angle and $\vec{u}$ the unit vector defining the direction of $\vec{S}$; and thus $f_i(\vec{S})f_j(\vec{S})\cos(\vec{S}.\vec{r}_{ij})$ is half of the amount of intensity scattered by the $ij$ pair of atoms, the other half scattered by this pair being $f_j(\vec{S})f_i(\vec{S})\cos(\vec{S}.\vec{r}_{ji})$] and its angularly averaged version, and known as

2) the DSE, the Debye formula etc., e.g. see [6]-[10]:

$$I(S) = \sum_{i=1}^{N}\sum_{j=1}^{N} f_i(S)f_j(S)\frac{\sin Sr_{ij}}{Sr_{ij}}. \quad (2)$$

where $S = |\vec{S}|$ and $r_{ij} = |\vec{r}_{ij}|$. To allow Fourier inversion of experimental X-rays, neutron or electron data from disordered materials, i.e. to get either of the so named radial distribution and pair correlation functions, this (2) is approximated to enable applying FTT, thus altering its value and, therefore, making the obtained real space results difficult to interpret, which is a quite serious problem. The here provided new expression relative to $I(S)$ as given by (2) allows avoiding this problem: its application to perform Fourier inversions is immediate and natural, i.e. does not need to resort to approximations at all.

To explicitly show the presence within the scattering volume (i) of $L$ different chemical species and (ii) of atom pairs which scatter with equal amounts of intensity, (1) and (2) can be rewritten to be given by, respectively, (3) {in which $I_m(\vec{S})$ and $I_{mk}(\vec{S})$ can be re-expressed, as is done here, to be given, respectively, by (4) and (5)} and by other possible equations, like (15) [obtainable from (3) through angularly averaging $I_m(\vec{S})$ and $I_{mk}(\vec{S})$ over the $4\pi$ solid angle, as done by Debye, [7], to get (2) from (1)].

### 2.1.1. Expression of $I(\vec{S})$ showing the presence of various chemical elements

This expression is known and is the following:

$$I(\vec{S}) = \sum_{m=1}^{L} I_m(\vec{S}) + 2 \sum_{m=1}^{L} \sum_{k>m}^{L} I_{mk}(\vec{S}), \qquad (3)$$

where $I_m(\vec{S})$ and $I_{mk}(\vec{S})$

a) represent: $I_m(\vec{S})$ the scattering by the pairs of atoms of the chemical species $m$ acting as if the atoms of the remaining chemical species were absent and $I_{mk}(\vec{S})$ the scattering by atom pairs involving atoms with different chemical natures, i.e. the $m$ and the $k$ species; and

b) can validly be rewritten, as is done here in the next paragraph, to have *more advantageous shapes but with exactly the same values*.

## 2.2. Continuing the work on $I_m(\vec{S})$ and $I_{mk}(\vec{S})$ to get superior expressions

### 2.2.1. Working on $I_m(\vec{S})$ to express it as (4)

The additional work done on $I_m(\vec{S})$ consists in grouping the terms $f_i(\vec{S})f_j(\vec{S})\cos(\vec{S}.\vec{r}_{ij})$ which are equal to each other, {so those with $f_i(\vec{S}) = f_j(\vec{S}) = f_m(\vec{S})$ and with associated interatomic vectors equal to one another plus equal to, e.g., $\vec{r}_{q_m}$}, to constitute the various components $n_m(\vec{r}_{q_m})$, $q_m = 0, 1, 2, \cdots$, of the distribution of the amounts of intensity $F_{q_m}(\vec{S}) \cos(\vec{S}.\vec{r}_{q_m}) = f_{q_m}^2(\vec{S}) \cos(\vec{S}.\vec{r}_{q_m}) = f_m^2(\vec{S}) \cos(\vec{S}.\vec{r}_{q_m})$ each of which is, therefore, associated with a type of interatomic vector, $\vec{r}_{q_m}$ typifying all pairs of atoms, present within the irradiated specimen volume, whose atoms are of the same chemical nature and are linked to one another by vectors equal to $\vec{r}_{q_m}$.

$$I_m(\vec{S}) = \sum_{q_m=0}^{M_m-1} n_m(\vec{r}_{q_m}) F_{q_m}(\vec{S}) \cos(\vec{S}.\vec{r}_{q_m}), \qquad (4)$$

where $M_m$ is an odd number and is the total number of the different components making up this (4); $n_m(\vec{r}_{q_m})$, $q_m = 0, 1, \cdots, (M_m - 1)/2, (M_m - 1)/2 + 1, (M_m - 1)/2 + 2, \cdots, M_m - 1$, is, therefore, the $m$ species-related sub part of the global DIV present within the irradiated specimen volume;[1] the component $n_m(\vec{r}_{q_m=0} = \vec{0})$ corresponds to the self pairing of the $m$ atoms and therefore is equal to the total number, $N_m$, of atoms of the $m$th chemical species present within the irradiated specimen volume; thus $n_m(\vec{r}_{q_m=0} = \vec{0}) = N_m$, whereas the remaining components of this sub distribution occur in pairs each of which is associated with a vector type, $\vec{r}_{q_m}$, and its opposite $-\vec{r}_{q_m}$. This is for the reason that each of the atom pairs $ij$ carries two vectors, $\vec{r}_{ij}$ and $\vec{r}_{ji}$, equal in length and opposite in direction; and $F_{q_m}(\vec{S}) = f_{q_m,1}(\vec{S})f_{q_m,2}(\vec{S})$ where $f_{q_m,1}(\vec{S})$ and $f_{q_m,2}(\vec{S})$ are the scattering factors of, respectively, the two atoms $q_{m,1}$ and $q_{m,2}$, both are of chemical nature $m$, making up the pair type $q_m$ and, thus, are both equal to the scattering factor of an atom of the species $m$, i.e. $f_{q_m,1}(\vec{S}) = f_{q_m,2}(\vec{S}) = f_m(\vec{S})$ and thus $F_{q_m}(\vec{S}) = f_m^2(\vec{S})$. A way of arranging the various components of (4) can be that, assumed here at this level, in which: $n_m(\vec{r}_{q_m=0}) = N_m$ corresponds to the component for which $\vec{r}_{q_m=0} = \vec{0}$ and $n_m(\vec{r}_{q_m=[q'_m+(M_m-1)/2]}) = n_m(\vec{r}_{q'_m})$ correspond to the components for which $\vec{r}_{q_m=[q'_m+(M_m-1)/2]} = -\vec{r}_{q'_m}$ where $q'_m = 1, 2, \cdots . (M_m - 1)/2$; the form of (4), on the other hand, makes calculating partial intensity distributions, $I_m(\vec{S})$, much easier; this is by calculating the partial DIVs and then inserting them into (4) ; and

### 2.2.2. Working on $I_{mk}(\vec{S})$ to express it as (5)

$I_{mk}(\vec{S})$ is reshaped through grouping the terms $f_i(\vec{S})f_j(\vec{S})\cos(\vec{S}.\vec{r}_{ij})$ [with $f_i(\vec{S}) \neq f_j(\vec{S})$] which are equal to one another (so whose respective interatomic vectors are equal to each other and, therefore, are all of the same type, e.g. written as $\vec{r}_{t_{mk}}$), to form the various, $t_{mk} = 1, 2, \cdots$, components $n_{mk}(\vec{r}_{t_{mk}})$ of the distribution of the amounts of intensity $F_{t_{mk}}(\vec{S})\cos(\vec{S}.\vec{r}_{t_{mk}}) = f_m(\vec{S})f_k(\vec{S})\cos(\vec{S}.\vec{r}_{t_{mk}})$ each of which is associated with an interatomic vector type, $\vec{r}_{t_{mk}}$, typifying all pairs of atoms, present within the scattering volume, whose atoms are of chemical natures $m$ and $k$ and are linked to one another by interatomic vectors equal to the vector type $\vec{r}_{t_{mk}}$ to be given by:

---

[1] This is for the reason that each of the interatomic vectors relative to the atom pairs made up of atoms of the same chemical nature contributes to the scattered intensity $I_m(\vec{S})$ with an amount of intensity represented by exactly one term $F_{q_m}(\vec{S}) \cos(\vec{S}.\vec{r}_{q_m}) = f_m^2(\vec{S}) \cos(\vec{S}.\vec{r}_{q_m})$.

$$I_{mk}(\vec{S}) = \sum_{t_{mk}=1}^{M_{mk}} n_{mk}(\vec{r}_{t_{mk}}) F_{t_{mk}}(\vec{S}) \cos(\vec{S}.\vec{r}_{t_{mk}}) \qquad (5)$$

where $M_{mk}$ is an even number and is the total number of the different components making up this (5); the distribution $n_{mk}(\vec{r}_{t_{mk}})$, $[t_{mk} = 1, 2, \cdots, M_{mk}/2, M_{mk}/2 + 1, M_{mk}/2 + 2, \cdots, M_{mk}]$, is also a sub part of the global DIV relative to the irradiated specimen volume; in this case of sub distribution there is no component $n_{mk}(\vec{r}_{t_{mk}} = \vec{0})$ in (5) as pairs of atoms made up of atoms with different chemical natures that have zero interatomic distance, and thus with zero interatomic vectors, cannot exist; however, one could, if wished, assume that it would exist but with a value that is necessarily zero and write $n_{mk}(\vec{r}_{t_{mk}} = \vec{0}) = 0$ for $t_{mk} = 0$ with the necessary accompanying meaning that this component does not exist. Thereafter including this null component into (5) will not change its value. In this case of sub distribution also the constituting components occur in pairs each of which is associated with a vector type, $\vec{r}_{t_{mk}}$, and its opposite $-\vec{r}_{t_{mk}}$; and $F_{t_{mk}}(\vec{S}) = f_{t_{mk,1}}(\vec{S}) f_{t_{mk,2}}(\vec{S})$ where $f_{t_{mk,1}}(\vec{S})$ and $f_{t_{mk,2}}(\vec{S})$ are, respectively, the scattering factors of the two atoms $t_{mk,1}$ and $t_{mk,2}$, making up the $t_{mk}th$ type of pairs of atoms, i.e. a type of pairs whose constituting atoms are, one, of chemical nature $m$ and, the other one, of chemical nature $k$; this gives, for instance, $f_{t_{mk,1}}(\vec{S}) = f_m(\vec{S})$ and $f_{t_{mk,2}}(\vec{S}) = f_k(\vec{S})$ and therefore $F_{t_{mk}}(\vec{S}) = f_m(\vec{S}) f_k(\vec{S})$. As for (4), a way of arranging the various components of (5) can be that, assumed at this level, in which $n_m(\vec{r}_{t_{mk}=[t'_{mk}+M_{mk}/2]}) = n_m(\vec{r}_{t'_{mk}})$ correspond to the components for which $\vec{r}_{t_{mk}=[t'_{mk}+M_{mk}/2]} = -\vec{r}_{t'_{mk}}$ where $t'_{mk} = 1, 2, \cdots . M_{mk}/2$.

The expression (3) of the scattered intensity, with $I_m(\vec{S})$ and $I_{mk}(\vec{S})$ given by, respectively, (4) and (5), is *exactly* equal to (1), can be named the 'reshaped diffracted intensity (RDI)' and is much more friendly than (1), as it can be much more easily studied and used than (1).

### 2.2.3. The, global, distribution of interatomic vectors

The global distribution of the interatomic vectors $n(\vec{r}_p)$, written in accordance with the definition proposed in [23], where the symbol $p = 1, 2, \cdots, M = \sum_{m=1}^{L} M_m + \sum_{m=1}^{L}\sum_{k=1}^{L} M_{mk}$ is used to label all of the components making up this global distribution, is determined by the combination of the various partial distributions $n_m(\vec{r}_{q_m})$ and $n_{mk}(\vec{r}_{t_{mk}})$ and can be expressed in different ways; for instance, a way can be that in which the various components of the various $I_m(\vec{S})$ and $I_{mk}(\vec{S})$ are arranged serially [and without including the null component, $n_{mk}(\vec{r}_0) = 0$, $\vec{r}_0 = \vec{0}$] to form a list, like that given by (6)[2], in which the various $n_m(\vec{r}_{q_m})$ components come before the $n_{mk}(\vec{r}_{t_{mk}})$ components,

$$m, k = 1, 2, \cdots, L$$

$$p = 1, 2, \cdots, M = \sum_{m=1}^{L} M_m + \sum_{m=1}^{L} \sum_{k=1}^{L} M_{mk}.$$

if $m = k$ and $p \leq \sum_{m=1}^{L} M_m$:  (this is $m = k$ case)

$$\begin{cases} q_m = \begin{cases} p - 1 \text{ if } p \leq M_1 \text{ and } m = 1, \\ p - 1 - \sum_{s=1}^{m-1} M_s \text{ if } \sum_{s=1}^{m-1} M_s < p \leq \sum_{s=1}^{m} M_s \text{ and } m > 1, \end{cases} \\ \vec{r}_p = \vec{r}_{q_m}, \quad \text{[atoms of pair } p \text{ are of same chemical nature } m\text{]} \\ n(\vec{r}_p) = n_m(\vec{r}_{q_m}), \end{cases} \quad (6)$$

else if $L > 1$ and $k > m$ and $p > \sum_{m=1}^{L} M_m$:  (this is $m \neq k$ case)

$$\begin{cases} t_{mk} = p - \sum_{s=1}^{L} M_s - \sum_{r=1}^{m-1} \sum_{s=1}^{L} M_{rs} - \sum_{s=1}^{k-1} M_{ms} \text{ if} \\ \sum_{s=1}^{L} M_s + \sum_{r=1}^{m-1} \sum_{s=1}^{L} M_{rs} + \sum_{s=1}^{k-1} M_{ms} < p \leq \sum_{s=1}^{L} M_s + \sum_{r=1}^{m-1} \sum_{s=1}^{L} M_{rs} + \sum_{s=1}^{k} M_{ms}, \\ \vec{r}_p = \vec{r}_{t_{mk}}, \quad \text{[atoms of pair } p \text{ are of chemical natures } m \text{ and } k \text{ (with } m \neq k\text{)]} \\ n(\vec{r}_p) = 2n_{mk}(\vec{r}_{t_{mk}}), \end{cases}$$

{where, we note, $n(\vec{r}_{p=1}) = n_m(\vec{r}_{q_m=0} = \vec{0}) = N_m$ with $m = 1$ (i.e. for $p = 1$); and $n(\vec{r}_p) = n(\vec{r}_{q_m=0} = \vec{0}) = N_m$, with $m > 1$ for the values of $p$ given by $p = 1 + \sum_{s=1}^{m-1} M_s$} which is fully validly usable together with (7), for this (7) is fully equivalent to (3) and to (1), since obtainable on insertion of $n_m(\vec{r}_{q_m})$ and $2n_{mk}(\vec{r}_{t_{mk}})$ expressed in terms of $n(\vec{r}_p)$ into (3):

$$I(\vec{S}) = \sum_{p=1}^{M} F_p(\vec{S}) n(\vec{r}_p) \cos(\vec{S} \cdot \vec{r}_p). \quad (7)$$

---

[2] The data used to draw the DID curves of figures 1 and 2 were written serially, i.e. in a way similar to the arrangement sort of scheme related to the equation that results on angularly averaging (7) over the full $4\pi$ solid angle as done by Debye to get the equation named after him: the Debye formula.

Figure 1 shows the $n(|\vec{r}_p|)$ calculated using the appropriate coordinates of the atom positions of zinc (Zn) and sulfur (S) for a rod-like parallelepiped ZnS perfect crystal (with dimensions [2a,2a,20a], a=5.41Å, e.g. [23]) obtained using the equation that results on averaging (6) over the 4π solid angle plus the corresponding X-rays profile shown in the inset.

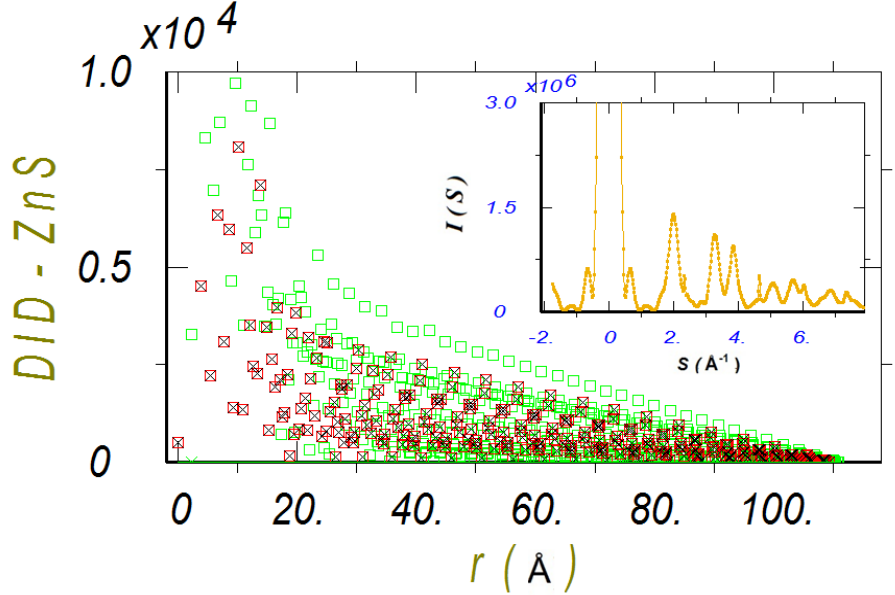

**Figure 1: Exact DID calculated from atom positions**. The two interatomic distances $|\vec{r}_{AB}|$ and $|\vec{r}_{BA}|$, whether $A \equiv B$ or not, corresponding to the two vectors $\vec{r}_{AB}$ and $\vec{r}_{BA}$ are considered as two equal but distinct interatomic distances, since corresponding to two distinct terms within the expression of the scattered intensity. Material stoichiometric ZnS crystal. ⊠: Zn-Zn and S-S partial DIDs; □: Zn-S partial DID; *Inset*: X-rays profile calculated using the plotted ZnS DID data. (We note the size and shape-related structure present at small angles of scattering. This is converted into diffuse background-like intensity by thermal agitation.)

### 2.3. Re-expressing (6) to take advantage of, two, particular facts

On the other hand, the expression (6) for $n(\vec{r}_p)$ can also be re-expressed taking advantage of the symmetry properties existing in connection with the **two facts** that **(a)** each atom pair corresponds to two interatomic vectors equal in length and opposite in direction and **(b)** in the particular case of crystals the presence of translational symmetry generates the existence of the possibility of doing simplifications.

#### 2.3.1. Taking advantage of the fact that two different interatomic vectors associate with each atom pair

Effectively, in general, on fact (a) $n_m(\vec{r}_{q_m})$ [with $q_m \neq 0$] and $n_{mk}(\vec{r}_{t_{mk}})$ involve the vectors $\vec{r}_{q_m}$ and $\vec{r}_{t_{mk}}$ and their respective opposites $-\vec{r}_{q_m}$ and $-\vec{r}_{t_{mk}}$, which allows giving (4) and (5) much more expressive, and useful, forms embodied, respectively, by the shapes of (8) and (10).

$$I_m(\vec{S}) = F_m(\vec{S}) \sum_{q'_m=-[M_m-1]/2}^{[M_m-1]/2} n'_m(\vec{r}_{q'_m}) \cos(\vec{S}.\vec{r}_{q'_m}), \tag{8}$$

where $n'_m(\vec{r}_{q'_m})$ has the following characteristics:

$$\begin{cases} q'_m = -\dfrac{(M_m-1)}{2}, \cdots, -2, -1, 0, 1, 2, \cdots, \dfrac{(M_m-1)}{2} \\ n'_m(\vec{r}_{q'_m}) = n_m(\vec{r}_{q'_m}) = N_m, \quad \text{if } q'_m = 0, \\ n'_m(\vec{r}_{q'_m}) = \dfrac{n_m(\vec{r}_{|q'_m|})}{2} \quad \text{if } q'_m \neq 0, \\ \text{so } n'_m(\vec{r}_{|q'_m|}) = n'_m(\vec{r}_{-|q'_m|}) \text{ and } \vec{r}_{|q'_m|} = -\vec{r}_{-|q'_m|}; \end{cases} \tag{9}$$

this (8) represents the partial DIV relative to the $m$th chemical species and is now even, symmetrical about the vertical axis. And

$$I_{mk}(\vec{S}) = F_{mk}(\vec{S}) \sum_{\substack{t'_{mk}=-M_{mk}/2 \\ t'_{mk} \neq 0}}^{M_{mk}/2} n'_{mk}(\vec{r}_{t'_{mk}}) \cos(\vec{S}.\vec{r}_{t'_{mk}}). \tag{10}$$

In this reshaped expression, (10), of $I_{mk}(\vec{S})$, the partial DIV $n'_{mk}(\vec{r}_{t'_{mk}})$ is even also and has the following characteristics:

$$\begin{cases} t'_{mk} = -\dfrac{M_{mk}}{2}, -\dfrac{M_{mk}}{2}+1, -\dfrac{M_{mk}}{2}+2, \cdots, -2, -1, 1, 2, \cdots, \dfrac{M_{mk}}{2} \\ n'_{mk}(\vec{r}_{t'_{mk}}) = \dfrac{n_{mk}(\vec{r}_{|t'_{mk}|})}{2}, \\ \text{so } n'_{mk}(\vec{r}_{|t'_{mk}|}) = n'_{mk}(\vec{r}_{-|t'_{mk}|}) \text{ and } \vec{r}_{|t'_{mk}|} = -\vec{r}_{-|t'_{mk}|}. \end{cases} \tag{11}$$

Therefore, the insertion of the two (8) and (10) into (3) produces a different, a reshaped further but not altered in value, expression for the diffracted intensity from a polyatomic material:

$$I(\vec{S}) = \sum_{m=1}^{L} F_m(\vec{S}) \sum_{q'_m=-[M_m-1]/2}^{[M_m-1]/2} n'_m(\vec{r}_{q'_m}) \cos(\vec{S}.\vec{r}_{q'_m})$$

$$+ 2 \sum_{m=1}^{L} \sum_{k>m}^{L} F_{mk}(\vec{S}) \sum_{\substack{t'_{mk}=-M_{mk}/2 \\ t'_{mk} \neq 0}}^{M_{mk}/2} n'_{mk}(\vec{r}_{t'_{mk}}) \cos(\vec{S}.\vec{r}_{t'_{mk}}) \quad (12)$$

that is entirely equivalent to the usual expression of the scattered intensity as given by (1), that is more advantageous[3] than (1) and that is the symmetrical expression of RDI. Additionally, this RDI for a polyatomic system can be re-expressed to make its evenness even more manifest. This is through suitably combining the various even partial distributions given by (9) and (11) to get an $n'(\vec{r}_{p'})$ that leads to $I(\vec{S})$ written as:

$$I(\vec{S}) = \sum_{m=1}^{L} F_m(\vec{S}) \sum_{p'=-(M-1)}^{M-1} n'(\vec{r}_{p'}) \cos(\vec{S}.\vec{r}_{p'}). \quad (13)$$

where $n'(\vec{r}_{p'})$ is given by:

$$\begin{aligned}
&m, k = 1, 2, \cdots, L, \\
&M = \sum_{m=1}^{L} M_m + \sum_{m=1}^{L} \sum_{k=1}^{L} M_{mk}. \\
&p' = -(M-1), -(M-2), \cdots, -2, -1, 0, 1, 2, \cdots, (M-2), (M-1), \\
&p = |p'| + 1 = 1, 2, \cdots, M, \\
&n'(\vec{r}_{p'}) = \begin{cases} n(\vec{r}_p), & \text{if } p' = 0 \text{ or } |p'| = \sum_{s=1}^{m-1} M_m, \\ \dfrac{n(\vec{r}_p)}{2}, & \text{if } p' \neq 0 \text{ and } |p'| \neq \sum_{s=1}^{m-1} M_m, \end{cases}
\end{aligned} \quad (14)$$

where $n(\vec{r}_p)$ is also given by (6). Under the form of (13), and particularly for the fundamental case of monatomic materials, $I(\vec{S})$ can directly, i.e. making no assumption and without altering its value, be given the shape of the very useful, and precisely known, discrete Fourier transform. This is therefore the form suitable for building Fourier transform-based analytical techniques which are capable of yielding experimental accurate results that are immediately and validly interpretable in terms of the physical concepts and principles underlying the kinematic approach to diffraction.

---

[3] This is especially true when it is about converting experimental reciprocal space data to real space data through the FTT, since this (12) can be used 'as it is' in connection with all material cases for which the kinematic approach to diffraction applies, i.e. without the need for modifying its value through assumption and approximation makings to permit the employment of the FTT.

Furthermore, in material system cases for which the interatomic vectors relevant to each interatomic vector type are distributed over the $4\pi$ solid angle with the randomness needed for averaging (12) and (13) as done by Debye (e.g. [2]) to get (2), one can get to two extra equations equivalent to each other and fully equivalent to (2) but with shapes much more favorable for constructing analytical techniques that can extract accurate experimental structural information embodies by DIDs from diffracted intensity profiles. One of these extra equations is obtained by angularly averaging (12) to get:

$$\langle I(\vec{S})\rangle = I(S) = \sum_{m=1}^{L} F_m(S) \sum_{q'_m=-[M_m-1]/2}^{[M_m-1]/2} n'_m\left(r_{q'_m}\right) \frac{\sin\left(Sr_{q'_m}\right)}{Sr_{q'_m}}$$

$$+ 2 \sum_{m=1}^{L} \sum_{k>m}^{L} F_{mk}(S) \sum_{\substack{t'_{mk}=-M_{mk}/2 \\ t'_{mk}\neq 0}}^{M_{mk}/2} n_{mk}(r_{t'_{mk}}) \frac{\sin(Sr_{t'_{mk}})}{Sr_{t'_{mk}}}; \qquad (15)$$

[where, therefore, $n'_m\left(r_{q'_m}\right)$ is the partial absolute, i.e. averaged relative to nothing, DID associated with the atoms of the chemical species $m$ acting as if they were totally isolated from the other chemical species atoms; $n_{mk}(r_{t'_{mk}})$ is the partial absolute DID relative to the interatomic distances involving one atom of chemical species $m$ and one atom of chemical species $k$]. This (15) is the *version* of the DSE which can produce, through the discrete Fourier transform, accurate structural data embodied by the DID from experimental intensity data at least for the fundamental case of monatomic material systems: in this case there is no need for considering extra assumptions or approximations; however, the determination of DIDs for polyatomic systems would necessitate more than one diffraction experiment, like when using anomalous X-ray scattering. Also, considered for monatomic systems, (15) reduces to somewhat resemble an expression for (2) recast, see [7], by Germer & White [9] in connection with randomly oriented 'face-centered cubic molecules'.

### 2.3.2. Taking advantage of the existence of translational symmetry in crystals

And on fact (b), in the particular case of crystals the existence of translational symmetry allows expressing $n'_m\left(\vec{r}_{q'_m}\right)$ and $n'_{mk}(\vec{r}_{t'_{mk}})$ using formulas, which is very useful indeed, as the re-expressing leads to valuable reshaped expressions for the DIV and, therefore, for the scattered intensity $I(\vec{S})$, e.g. as given by (12). Thus formulas giving DIVs can be obtained for parallelepiped crystals. Only the formula for monatomic crystal is given here, as the full description of, for instance just, the formula relative to the diatomic crystals case needs a bit of extra writing space.

### 2.3.3. Formulas for DIV, DID and $I(S)$ relative to monatomic crystals

o   The DIV formula. The following (16)

$$n'(\vec{r}_{u'v'w'}) = (N_x - |u'|)(N_y - |v'|)(N_z - |w'|)$$
$$\vec{r}_{u'v'w'} = u'\vec{a} + v'\vec{b} + w'\vec{c}$$
$$\begin{cases} u' = -(N_x - 1), \cdots, -1, 0, 1, \cdots, N_x - 1, \\ v' = -(N_y - 1), \cdots, -1, 0, 1, \cdots, N_y - 1, \\ w' = -(N_z - 1), \cdots, -1, 0, 1, \cdots, N_z - 1. \end{cases} \qquad (16)$$

{where $\vec{a}$, $\vec{b}$ and $\vec{c}$: fundamental vectors; $N_x$, $N_y$ and $N_z$: numbers of atoms along, respectively, x, y and z: axes; so dimensions, and therefore size and shape, of crystal defined by $(N_x - 1)|\vec{a}|$, $(N_y - 1)|\vec{b}|$ and $(N_z - 1)|\vec{c}|$. **Correspondence with (9):** $L = 1$ so $m = 1$, $n'_m(\vec{r}_{q'_m}) = n'(\vec{r}_{u'v'w'})$ so $q'_m$ split into the three $\vec{r}_{u'v'w'}$, interatomic vector type, determining $u'$, $v'$ and $w'$ parameters. **Remarks**: the interatomic vector $\vec{r}_{u'v'w'}$ type in (16) and the usual translational vector $\vec{T} = n_1\vec{a} + n_2\vec{b} + n_3\vec{c}$ have similar formulations, which means that the vectors defined by $\vec{r}_{u'v'w'}$ in (16), firstly, are equal to some of those corresponding to $\vec{T}$ and, secondly, some of them, those relative to $u' = 0, 1, \cdots, N_x - 1$, $v' = 0, 1, \cdots, N_y - 1$ and $w' = 0, 1, \cdots, N_z - 1$, are identical to the vectors giving the positions of all atoms of the crystal relative to an atom set at $\vec{r}_{000} = \vec{0}$ taken as origin.} is the DIV formula for the triclinic lattice type-related material systems obtained for parallelepiped monatomic crystals and thus leads, on insertion into (8) or (12), to the formula for the diffracted intensity from these crystals. The three $u'$, $v'$ and $w'$ DIV component-determining parameters connect with the Miller, and diffraction peak-defining, $h$, $k$ and $l$ indices through $(u'h + v'k + w'l) = n$, $n$ being an integer, zero included.

This is with the **note** that this formula (16) can be obtained through determining, getting, the numbers of the zero-, one-, two- and three-dimensional cells (j) defined by three edges vectors, e.g. written as $u\vec{a}$, $v\vec{b}$ and $w\vec{c}$ where the three $u$, $v$ and $w$ are integer parameters and $\vec{a}$, $\vec{b}$ and $\vec{c}$ are the three lattice base vectors, and (jj) symbolically represented by $(u,v,w)$. Thus, the ensemble of these cells is given

- partly by (0,0,0) cells, which thus correspond to simple lattice points and their number is equal to the lattice points within the monatomic crystal assumed,

- partly by ($u{\neq}0,0,0$), ($0,v{\neq}0,0$) and ($0,0,w{\neq}0$) cells, which thus correspond to the linear arrays of lattice points along, respectively, the [100], [010] and [001] directions,

- partly by ($u{\neq}0,v{\neq}0,0$), ($u{\neq}0,0,w{\neq}0$) and ($0,v{\neq}0,w{\neq}0$) cells, which therefore correspond to the planar arrays of lattice points parallel to, respectively, the x-y, z-x and y-z planes and

- partly by ($u{\neq}0,v{\neq}0,w{\neq}0$) cells, these are the usual three dimensional cells.

This is for the reason that the ensemble of the diagonals of these cells carries the ensemble of the interatomic vectors of the crystal. The diagonal of each of the (0,0,0) cells carries one null interatomic vector, i.e. $\vec{0}$; related to the atom self-pairing. Each of the diagonals of the ($u$,0,0), (0,$v$,0) and (0,0,$w$) cells carries two interatomic vectors equal in length and opposite in direction. Each of the ($u$,$v$,0), ($u$,0,$w$) and (0,$v$,$w$) cells has two diagonals each of which carries two interatomic vectors and thus has four interatomic vectors. And the ($u{\neq}0,v{\neq}0,w{\neq}0$) cells have four diagonals each and thus have eight interatomic vectors each. The numbers of the cell copies for each of the cell types ($u=0,v=0,w=0$), ($u{\neq}0,0,0$), ($0,v{\neq}0,0$), ($0,0,w{\neq}0$), ($u{\neq}0,v{\neq}0,0$), ($u{\neq}0,0,w{\neq}0$), ($0,v{\neq}0,w{\neq}0$) and ($u{\neq}0,v{\neq}0,w{\neq}0$) can all be written as $(N_x - u)(N_y - v)(N_z - w)$ where $N_x$, $N_y$ and $N_z$ are the numbers of the crystal atoms located on the edges along the x-, y- and z-axis, respectively, and $u = 0, 1, \cdots, N_x$, $v = 0, 1, \cdots, N_y - 1$ and $w = 0, 1, \cdots, N_z - 1$. But if one uses the parameters $u' = -(N_x - 1), \cdots, -1, 0, 1, \cdots, N_x - 1$, $v' = -(N_y - 1), \cdots, -1, 0, 1, \cdots, N_y - 1$ and $w' = -(N_z - 1), \cdots, -1, 0, 1, \cdots, N_z - 1$, then any interatomic vector type within the crystal can be written as $\vec{r}_{u'v'w'} = u'\vec{a} + v'\vec{b} + w'\vec{c}$ and found to be present within the crystal with $n(\vec{r}_{u'v'w'}) = (N_x - |u'|)(N_y - |v'|)(N_z - |w'|)$ copies. Hence (16).

o <u>The DID formula.</u> Moreover, if one takes the modulus of the vectors within this (16) then one gets the formula of the DID for parallelepiped monatomic crystals written as:

$$n(r_{u'v'w'}) = \begin{cases} N_x N_y N_z & \text{if } u' = v' = w' = 0, \\ 2(N_x - u')(N_y - v')(N_z - w'), & \text{otherwise,} \end{cases}$$

$$r_{u'v'w'} = |u'\vec{a} + v'\vec{b} + w'\vec{c}|,$$

$$\begin{cases} u' = 0, 1, \cdots, N_x - 1, \\ v' = 0, 1, \cdots, N_y - 1, \\ w' = 0, 1, \cdots, N_z - 1. \end{cases} \quad (17)$$

or written in the way which allows inserting it into (15) to get the intensity diffracted by, for instance, "a parallelepiped randomly moving crystal".

o  The intensity diffracted by, for instance, "a parallelepiped randomly moving crystal".

$$I(S) = f^2(S) \sum_{u'=-(N_x-1)}^{N_x-1} \sum_{v'=-(N_y-1)}^{N_y-1} \sum_{w'=-(N_z-1)}^{N_z-1} (N_x - |u'|)(N_y - |v'|)(N_z - |w'|) \frac{\sin\left(S|u'\vec{a} + v'\vec{b} + w'\vec{c}|\right)}{S|u'\vec{a} + v'\vec{b} + w'\vec{c}|}. \quad (18).$$

This (18) is, possibly, the simplest way of writing the formula for calculating intensities diffracted by monatomic crystals with parallelepiped shapes, but writing it in ways that give access to shorter calculation times is possible.

## 2.4. Calculating DIDs and intensity profile using formula: example and software

### 2.4.1. Example

(17) is a lot easier to use, than (16), when checking (for instance through comparison with results obtained by other ways, since DIDs can also be determined, example figure 1, from the knowledge of the atom positions within crystals) the exactitude of (16) is the aim sought. Figure 2 compares the two DID versions for a small parallelepiped polonium (Po) perfect crystal (edges $5|\vec{a}|$, $5|\vec{b}|$ and $10|\vec{c}|$, lattice = simple cubic, $|\vec{a}| = |\vec{b}| = |\vec{c}| = 3.34$ Å, 396 atoms) obtained one through the formula given by (17) and the other one from atom positions. There is no difference between the two curves, which signs (16) is exact. The X-rays profile of the Figure 2 inset was obtained using the shown DID and the values for Po-related X-rays atomic scattering factor parameters, taken from [25], usable together with the parameterized representation of atomic X-rays scattering factors due to Doyle and Turner, [26]. This profile, also, shows a structure, between the zero angle of scattering and the first diffraction peak, associated with the shape and size of the crystal considered. This structure is converted into diffuse background-like intensity by thermal agitation. The code of the program used to get the data of Figure 2 is easily reachable, [27]. On the other hand, the calculations through formulas are much more rapidly done than from atom positions. For example, the DID for a monatomic simple cubic crystal with 1000000 atoms can be calculated using an expanded version of (17), such as that used to build 'option "F"' of DIDprofile01, see [27], within minutes, i.e. less than 10 minutes using a portable computer with moderate speed and would take days or even weeks using the same computer.

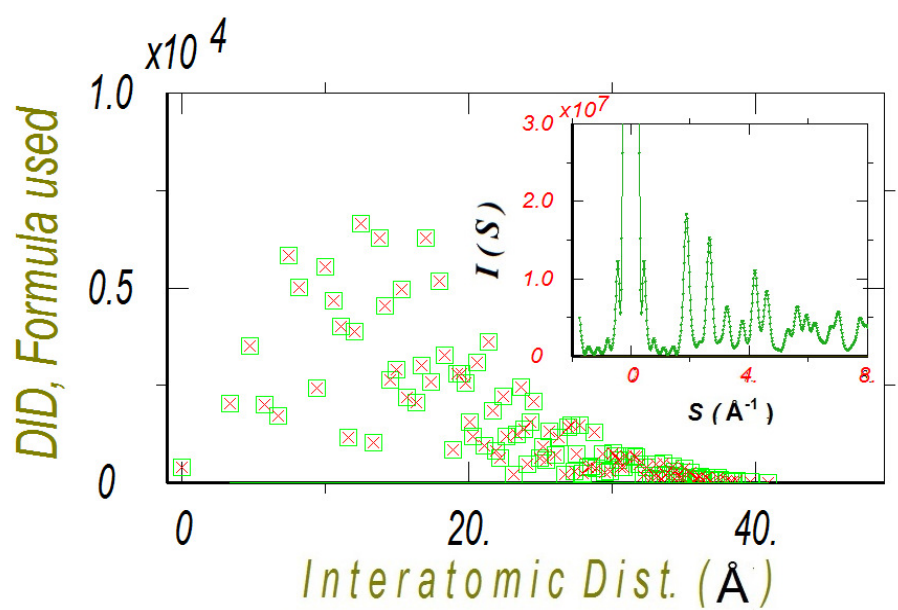

Figure 2: **Comparing two DIDs for a parallelepiped crystal of polonium.** One, squares, calculated using formula given by (17) and the other one, crosses, calculated from atom positions. The two DIDs are identical. *Inset*: X-ray profile calculated through (15) using DID data obtained through formula.

### 2.4.2. Software

The data used to draw the curves of Figure 2 can be obtained through the program "DIDprofile01" which results from compiling and linking the code source "Calculate_Distribution_Interatomic_Distances_And_Diffr_Profile_Monat_Cryst_Using_Formulas_01.f90", [27]. These data can be obtained thought three different routines each of which corresponds to an option within the Menu of the DIDprofile01. The routines were written on the bases of three algorithms. They were built using 1) the compact formula represented by (16) without altering its shape, 2) after writing it to consist of sub-formulas corresponding to the sub-DIDs relative to the various linear directions possible for the parallelepiped crystal under consideration and 3) after additional efforts to reach a much faster application, see https://osf.io/axuym/wiki/home/ for more details.

## 3. Conclusion

Progressing further with the theoretical background underlying the usually employed means for acquiring knowledge through diffraction valuable developments have been reached: the exact scattered intensity was obtained with the form which, for instance, makes employing the kinematical diffraction theory together with the Fourier transform tool with no additional approximation and no additional assumption to get real space experimental data possible. Analytical techniques that (i) include intensity calculations either through formulas or from known atom positions and (ii) use the Fourier transform tool to produce, for both ordered and disordered atom arrangements, accurate experimental real-space data from reciprocal-space data which are *immediately and validly interpretable on the basis of the physical concepts and principles the kinematic approach to diffraction relies on have become possible*. The first ever produced formulas for calculating the distribution of the interatomic vectors and the distribution of the interatomic distances relative to a parallelepiped monatomic triclinic crystal are made available here together with a possible formula for calculating intensities diffracted by monatomic crystals. Plot examples of "distributions of interatomic distances"- and diffracted intensity-related data calculated both through formulas and from atom positions are given. The validation of the formulas provided was through comparing data calculated through formula and from atom positions.